\documentclass[preprint,showpacs,amsmath,amssymb,prl]{revtex4-1}

\usepackage{amssymb}
\usepackage{amsfonts}
\usepackage{amsmath}
\usepackage{graphicx}
\usepackage{dcolumn}
\usepackage{bm}
\usepackage{textcomp}
\usepackage{epstopdf}
\usepackage{hyperref}
\usepackage{wasysym}
\usepackage{xcolor}

\begin{document}


\title{Non-local opto-electrical spin injection and detection in germanium at room temperature} 

\author{F. Rortais$^{1}$, C. Zucchetti$^{2}$, L. Ghirardini$^{2}$, A. Ferrari$^{1}$, C. Vergnaud$^{1}$, J. Widiez$^{3,4}$, A. Marty$^{1}$, J.-P. Attan\'e$^{1}$, H. Jaffr\`es$^{5}$, J.-M. George$^{5}$, M. Celebrano$^{2}$, G. Isella$^{2}$, F. Ciccacci$^{2}$, M. Finazzi$^{2}$, F. Bottegoni$^{2}$ and M. Jamet$^{1*}$}
\affiliation{$^{1}$Spintec, Institut Nanosciences et Cryog\'enie, Univ. Grenoble Alpes, CEA, CNRS, F-38000 Grenoble, France\\
$^{2}$LNESS-Dipartimento di Fisica, Politecnico di Milano, 20133 Milano, Italy\\
$^{3}$Univ. Grenoble Alpes, F-38000 Grenoble, France\\
$^{4}$CEA, LETI, MINATEC Campus, F-38054 Grenoble, France\\
$^{5}$Unit\'e Mixte de Physique, CNRS, Thales, Univ. Paris-Sud, Univ. Paris-Saclay, 91767 Palaiseau, France
}%


\maketitle

\textbf{Non-local carrier injection/detection schemes lie at the very foundation of information manipulation in integrated systems. This paradigm consists in controlling with an external signal the channel where charge carriers flow between a "source" and a well separated "drain". The next generation electronics may operate on the spin of carriers instead of their charge \cite{Awschalom2007,Zutic2004} and germanium appears as the best hosting material to develop such a platform for its compatibility with mainstream silicon technology and the long electron spin lifetime at room temperature \cite{Li2012}. Moreover, the energy proximity between the direct and indirect bandgaps \cite{Pezzoli2012} allows for optical spin injection and detection within the telecommunication window \cite{Suess2013}. In this letter, we demonstrate injection of pure spin currents (\textit{i.e.} with no associated transport of electric charges) in germanium, combined with non-local spin detection blocks at room temperature. Spin injection is performed either electrically through a magnetic tunnel junction (MTJ) or optically, exploiting the ability of lithographed nanostructures to manipulate the distribution of circularly-polarized light in the semiconductor. Pure spin current detection is achieved using either a MTJ or the inverse spin-Hall effect (ISHE) across a platinum stripe. These results broaden the palette of tools available for the realization of opto-spintronic devices.}

Spintronics aims at exploiting the spin degree of freedom to manipulate information, while in conventional electronics, information is associated with the charge of carriers \cite{Wolf2001}. In this regard, $n$-type germanium appears as the best hosting material for spin transport and manipulation. The electron spin lifetime can reach several nanoseconds at room temperature \cite{Li2012} and the compatibility with mainstream silicon technology allows exploiting the spin-related properties of low dimensional SiGe-heterostructures \cite{Bottegoni2011}. Electrical spin injection and detection has been explored in Ge films or nanowires using either non-local measurements in lateral or vertical spin valves \cite{Zhou2011,Chang2013,Liu2010,Li2013} or the Hanle effect in three-terminal devices \cite{Jain2011,Saito2011,Rortais2016,Jain2012a,Jain2012b,Jeon2011,Hanbicki2011,Iba2012}. So far, the non-local lateral geometry is the most interesting one for the development of spintronics since the spin can be manipulated in the Ge channel between the spin injector and detector. However, experimental measurements have been limited in temperature to 225 K \cite{Zhou2011} and the only demonstration at room temperature used an indirect method based on the combination of spin pumping and inverse spin Hall effect (ISHE) \cite{Dushenko2015}.\\
Another important feature of bulk Ge is related to the energy proximity between its direct and indirect bandgaps \cite{Pezzoli2012}. This "quasi-direct" band structure was previously used to investigate spin dynamics in bulk Ge \cite{Bottegoni2013,Bottegoni2015} and SiGe nanostructures \cite{Guite2011}. It also gives the opportunity to develop opto-spintronic devices where photons can generate and detect spin currents. Here, we implement novel non-local spin injection/detection building blocks in germanium at room temperature, adding new functionalities to the common architectures available for spintronic devices. We demonstrate the lateral spin transport in $n$-type germanium-on-insulator (GeOI, $n$=2$\times$10$^{19}$ cm$^{-3}$) and a lightly $n$-doped bulk Ge sample ($n$=1.7$\times$10$^{16}$ cm$^{-3}$), exploiting electrical and optical spin generation respectively. The non-local spin detection is achieved using a MTJ or the ISHE in a Pt bar. By this, we can accurately extract the spin diffusion length $l_{sf}$ in the GeOI sample and the Ge substrate at room temperature, obtaining $l_{sf}$=650$\pm$30 nm and $l_{sf}$=12.4$\pm$0.4 $\mu$m, respectively. We also show direct optical mapping of spin diffusion in Ge and, by combining optical spin orientation and the ISHE in Pt, we build a non-local spin injection/detection scheme without the use of any ferromagnetic metal.\\
For electrical spin injection and detection, we use lateral spin valves (LSVs) fabricated on GeOI. The Ge layer is 1 $\mu$m-thick with uniform $n$-type heavy doping ($n$=2$\times$10$^{19}$ cm$^{-3}$) to favor electrical conduction and reduce the width of the Schottky barrier (see \textit{Methods} for details). The SiO$_{2}$ buried oxide layer (BOX) is also 1$\mu$m-thick. Electrical spin injection and detection are achieved using a MgO-based MTJ to avoid the impedance mismatch issue \cite{Fert2001}. Moreover, in order to reduce the density of localized states at the MgO/Ge interface \cite{Jain2012a,Tran2009}, we have grown the magnetic tunnel junction Pd(5nm)/Fe(15nm)/MgO(2.5nm) by epitaxy on Ge(100). The overall epitaxial relationship is Fe[100]$\vert\vert$MgO[110]$\vert\vert$Ge[100] as illustrated by the RHEED patterns along the [110] and [100] crystal axes of Ge in Fig.~\ref{Fig1}a. The sample is then processed into LSVs made of two magnetic tunnnel junctions and two ohmic contacts as schematically shown in Fig.~\ref{Fig1}b and detailed in \textit{Methods}. An example of LSV is shown in Fig.~\ref{Fig1}c where the gap, defined as the distance between the ferromagnetic electrodes edges, is 0.5 $\mu$m. The soft and hard magnetic electrodes have been processed along the [100] crystal axis of Fe and their dimensions are 1$\times$20 $\mu$m$^{2}$ and 0.5$\times$20 $\mu$m$^{2}$ respectively. The MTJs $I(V)$ curves are almost linear and their resistance-area product is 430 $\Omega \mu$m$^{2}$. Electrical measurements are performed in 3 different geometries as depicted in Fig.~\ref{Fig2} using a 2 $\mu$m gap between the two MTJs. The applied DC electrical current is 10 mA. In Fig.~\ref{Fig2}a and \ref{Fig2}b, the magnetic field is applied at 45$^{\circ}$ from the Fe electrode axis \textit{i.e.} along the [110] crystal direction which corresponds to the hard axis. This geometry allows to enhance and better separate the switching fields of the ferromagnetic electrodes in order to stabilize the antiparallel state.\\
In Fig.~\ref{Fig2}a, the DC electrical current is applied between the two ferromagnetic electrodes and the voltage is measured on the same contacts. This configuration is called the local or GMR (for giant magnetoresistance) configuration. The resistance is minimum when the Fe electrode magnetizations are parallel ($R_{\uparrow\uparrow}$) and maximum when they are antiparallel ($R_{\uparrow\downarrow}$). We find $\Delta R_{GMR}=R_{\uparrow\downarrow}-R_{\uparrow\uparrow}\approx$45 m$\Omega$.\\
The configuration of Fig.~\ref{Fig2}b is the non-local (NL) geometry: the current is applied between one pair of ferromagnetic-ohmic contacts and the voltage measured on the other pair. By this, only a pure spin current flows between the two ferromagnets without any charge current avoiding the contribution from spurious tunneling magnetoresistance effects to the detected signal \cite{Song2014,Txoperena2014}. The measured magnetoresistance signal $\Delta R_{NL}=R_{\uparrow\downarrow}-R_{\uparrow\uparrow}\approx$-25 m$\Omega$ is approximately half the GMR signal in amplitude and is independent on the applied bias voltage (not shown). It can be written as \cite{Jedema2002}: $\Delta R_{NL}=-\frac{P_{I}P_{D}}{\sigma A}l_{sf}exp\left(-\frac{L}{l_{sf}}\right)$. $P_{I}$ (resp. $P_{D}$) is the spin polarization of the tunnel current at the injection (resp. detection) electrode, $\sigma$ and $A$ are the conductivity and cross-sectional area of the Ge channel respectively. $L$ is the distance between the two ferromagnetic electrodes and $l_{sf}$ the spin diffusion length in Ge.\\
In the last geometry of Fig.~\ref{Fig2}c, the magnetic field is applied perpendicular to the Ge film and we measure the Hanle effect. In this configuration, injected spins experience the Larmor precession and the spin signal $\Delta R_{NL}^{Hanle}$ decays following roughly a Lorentzian curve. Starting from the parallel state, the spin signal writes \cite{Jedema2002}: 

\begin{equation}
\Delta R_{NL}^{Hanle}(B_{\bot})=\frac{P_{I}P_{D}D}{\sigma A}\int_{0}^{\infty}P(t)cos(\omega_{L}t)exp(-\frac{t}{\tau_{sf}})dt
\label{Hanle}
\end{equation}

where $D$ is the electron diffusion coefficient we determined independently using double Hall crosses at room temperature: $D$=23.5 cm$^{2}$s$^{-1}$. $P(t)=[1/\sqrt{4\pi Dt}]exp[-L^{2}/(4Dt)]$, $\omega_{L}=g\mu_{B}B_{\bot}/\hbar$ is the Larmor angular frequency with $g$=1.6 the $g$-factor of electrons in the Ge conduction band \cite{Feher1959}, $\mu_{B}$ the Bohr magneton and $\hbar$ the reduced Planck's constant. $\tau_{sf}$ is the electron spin lifetime in Ge. By fitting the Hanle curve in Fig.~\ref{Fig2}c, we find: $P_{I}P_{D}=$0.28$\pm$0.05 and $l_{sf}=$650$\pm$30 nm corresponding to a spin lifetime $\tau_{sf}=$180$\pm$20 ps. The spin diffusion length is in very good agreement with that obtained by Dushenko \textit{et al.} \cite{Dushenko2015} who found $l_{sf}$=660$\pm$200 nm at room temperature in a Ge film with comparable $n$-type doping by non-local spin pumping. This value is also in good agreement with that obtained by three-terminal measurements at room temperature: 1300 nm for $n$=10$^{18}$ cm$^{-3}$ \cite{Jain2012a} and 530 nm for $n$=1.0$\times$10$^{19}$ cm$^{-3}$ \cite{Jeon2011}.\\
This agreement between non-local and three-terminal Hanle measurements can be attributed to the fact that interface states are only weakly confined at room temperature and have little influence on the spin injection mechanism \cite{Jain2012a}. Finally, the GMR and NL magnetoresistance signals in Fig.~\ref{Fig2} both exhibit an overall Lorentzian shape around $\mu_{0}$H=0 T. This is due to the unavoidable oblique Hanle effect since the magnetic field is applied at 45$^{\circ}$ from the injected spin direction. This interpretation is supported by the fact that the full width at half maximum (FWHM) of the oblique Hanle curves in Fig.~\ref{Fig2}a and \ref{Fig2}b is exactly $\sqrt{2}$ times the FWHM of the Hanle curve of Fig.~\ref{Fig2}c where the applied field is perpendicular to the injected spins.


As a further step, we implement an optical spin injection block, combined with non-local lateral spin detection using a MTJ or a Pt stripe to detect the ISHE. In this case, a net electron spin polarization is generated by optical spin orientation \cite{Optical1984}. In this process, the absorption of circularly-polarized light generates spin polarized electron-hole pairs at the $\Gamma$ point of the Brillouin zone. The spin polarization of photogenerated electrons in the conduction band is $P=\left(n_{\uparrow}-n_{\downarrow}\right)/\left(n_{\uparrow}+n_{\downarrow}\right)$, being $n_{\uparrow(\downarrow)}$ the up- (down-) spin densities referred to the quantization axis given by the direction of light propagation in the material. Photogenerated holes are rapidly depolarized due to their very short spin lifetime \cite{Rortais2016}. If the incident photon energy is tuned to the direct Ge bandgap, an electron spin polarization $P = 50\%$ can be achieved \cite{Optical1984}. Right after the photogeneration, spin-oriented electrons thermalize from the $\Gamma$ to the L-valleys within $\approx 300$ fs, maintaining most of their spin polarization \cite{Pezzoli2012}. Here, we use a diffraction-limited confocal setup shown in Fig.~\ref{Fig3}a. The Ge(001) substrate is 450-$\mu$m-thick and lightly As-doped ($n$=1.7$\times$10$^{16}$ cm$^{-3}$). At normal incidence, only an out-of-plane spin polarization is generated, preventing any electrical spin detection in Ge with in-plane MTJs or the ISHE in Pt stripes. We circumvent this limitation by patterning Pt or Pt/MgO nanostructured stripes on the Ge substrate \cite{Bottegoni2014}. When the sample is illuminated with circularly-polarized light, the metallic pattern introduces a modulation of the amplitude and phase of the incoming electromagnetic wavefront. As shown in Fig.~\ref{Fig3}b and \ref{Fig3}c, the absorption of the scattered light in the Ge substrate results in the generation of complementary opposite electron spin populations inside Ge in correspondence to the edges of Pt or Pt/MgO nanostructures, with a net spin polarization lying in the plane of the device \cite{Bottegoni2014}. The resulting spin accumulation at each edge of a nanostructure creates a pure spin current detected non-locally by an in-plane magnetized MTJ or by ISHE in a Pt bar. As an example, the spin detection with a MTJ is detailed in Fig.~\ref{Fig3}d. The two non-local devices are depicted in Fig.~\ref{Fig4}a and \ref{Fig4}b respectively. They consist of a MTJ Pt(5nm)/Fe(15nm)/MgO(3.5nm) with dimensions 2$\times$2 $\mu$m$^{2}$ [resp. a Pt(15nm) bar with dimensions 3$\times$1 $\mu$m$^{2}$] acting as a spin detector and eight stripes of Pt(15nm)/MgO(8nm) [resp. Pt(15nm)] with dimensions 1$\times$2 $\mu$m$^{2}$ and separated by 1 $\mu$m (see \textit{Methods}). For the MTJ device, the MgO layer separating Pt from the Ge substrate is meant to avoid spin absorption in the Pt layer whereas, for the detection block based on the Pt bar, spin absorption by the Pt layer is necessary to detect the ISHE signal and Pt is in direct contact with the Ge surface. Scanning electron microscopy images of the final devices are shown in Fig.~\ref{Fig4}c for the MTJ device and Fig.~\ref{Fig4}d for the ISHE device. The measurement details are given in the \textit{Methods}. The optical images of the nanostructures are shown in Fig.~\ref{Fig5}a and \ref{Fig5}b. The voltage signals are recorded at the same time as the optical images at room temperature. They are reported in Fig.~\ref{Fig5}c for the MTJ (incident power $W=60$ $\mu$W) and in Fig.~\ref{Fig5}d for the Pt ISHE pad (incident power $W=1.8$ mW) respectively. In Fig.~\ref{Fig5}e and \ref{Fig5}f, the average across the Pt stripes of the MTJ and ISHE voltage signals are reported as a function of the distance to the detector for the MTJ and ISHE devices respectively. We clearly see an alternating signal that decays when the illuminating beam is moved away from the detector as a consequence of the finite spin diffusion length in Ge. As shown in Fig.~\ref{Fig5}e and \ref{Fig5}f, the voltage $\Delta V$ normalized to the light power $W$ can be easily fitted using the following expression: $\Delta V/W\propto sin(\omega x)\times exp(-x/l_{sf})$. $\omega$=2$\pi$/$L$ where $L$=2 $\mu$m is the pattern periodicity and $l_{sf}$ is the spin diffusion length. $x$ and $l_{sf}$ are in $\mu$m and $x$=0 corresponds to the position of the detector. By using such simple expression, we assume a one-dimensional spin diffusion model which is a rough approximation considering the three-dimensional geometry of our system. However, the very good agreement between the fitting curve and the experimental data suggests that the spin diffusion mostly takes place along $x$ which is probably due to the partial spin absorption by the Pt bars. We find: $l_{sf}$=12.4$\pm$0.4 $\mu$m for the MTJ device and $l_{sf}$=8$\pm$2 $\mu$m for the ISHE device. The difference between these values is related to the different sample geometry: in the MTJ device, a MgO layer separates Pt bars and Ge, whereas in the other one, Pt is directly in contact with Ge. Platinum acts as a spin sink: the presence of Pt bars between the generation and detection points reduces the number of spins reaching the detector. In the MTJ device, MgO prevents this mechanism and the exponential decay can be mostly related to depolarization in the semiconductor. On the contrary, for ISHE detection, spin absorption in the Pt pads cannot be neglected and this reduces the effective spin diffusion length as well as the detected electrical signal. To confirm this picture we fabricated a test ISHE device with MgO separating Pt and Ge, thus lowering the Pt spin-absorption, \textit{i.e.} the ISHE detector efficiency. As expected, this dramatically suppressed the detected signal. Using an upper bound of 100 cm$^{2}$s$^{-1}$ for the diffusion coefficient of electrons in the substrate\cite{Ioffe}, we obtain a spin lifetime $\tau_{sf}\approx$12 ns which is longer than the value predicted by Li \textit{et al.}\cite{Li2012}. However, the presence of Pt stripes might modify the atomic and electronic structure at the Pt/Ge interface where the spin transport takes place. Finally, the combination of optical spin orientation with in-plane polarization and ISHE in a Pt bar defines an original non-local spin injection/detection scheme without the use of any ferromagnetic metal which represents a new paradigm in the field of semiconductor spintronics.\\
In summary, we have demonstrated spin transport in Ge at room temperature using non-local electrical detection. Two different systems have been investigated: a heavily $n$-doped GeOI for electrical spin injection using a magnetic tunnel junction and a high quality lightly doped Ge substrate for optical spin orientation. In GeOI, we used a second MTJ for electrical spin detection in a lateral spin valve geometry whereas we used both a MTJ and the inverse spin Hall effect in a Pt stripe to detect the spin signal in the Ge substrate. In the latter case, we are only sensitive to in-plane spin polarization which is achieved by optical spin orientation at the edge of Pt nanostructures grown on Ge. In GeOI, we find a spin diffusion length $l_{sf}$=650$\pm$30 nm in perfect agreement with previous estimations at room temperature. In the Ge substrate, we find: $l_{sf}$=12.4$\pm$0.4 $\mu$m and only 8$\pm$2 $\mu$m when the Pt stripes partly absorb the spin current. Finally, we conclude that optical spin orientation in a confocal microscope with in-plane spin polarization is a very powerful tool to directly image the spin diffusion in Ge and more generally in any direct bandgap semiconductors.

\section*{Methods}

\subsection*{Electrical spin injection and detection}
GeOI was used to well-define the conduction channel and it was fabricated using the Smart Cut$^{TM}$ process from Ge epitaxially grown on Si at low temperature (400$^{\circ}$C) \cite{Reboud2016}. By using short duration thermal cycling under H$_{2}$ atmosphere, the threading dislocation density was reduced down to 10$^{7}$ cm$^{-2}$. However, a residual tensile strain of +0.148 \% (as determined by grazing incidence x-ray diffraction) built up during the cooling-down to room temperature after the thermal cycling due to the difference of thermal expansion coefficients between Ge and Si. The Ge layer is protected against oxidation by a 10 nm-thick SiO$_{2}$ film which is removed using hydrofluoric acid before the introduction into the molecular beam epitaxy (MBE) chamber. The native Ge oxide top layer was then thermally removed by annealing under ultrahigh vacuum. After this cleaning procedure, the reflection high-energy electron diffraction (RHEED) pattern exhibited a well-defined and high-quality (2$\times$1) surface reconstruction as the one shown in Fig.~\ref{Fig1}a. The nanofabrication of LSVs required 5 successive electron beam lithography levels and the key steps are: (i) the ion beam etching of the ferromagnetic electrodes using metallic hard masks, (ii) the growth of ohmic contacts made of Au(250nm)/Ti(10nm) by electron beam evaporation and lift-off technique and (iii) the deposition of a 100 nm-thick SiO$_{2}$ passivation layer by ion beam deposition (IBD) to insulate the bonding pads from the Ge channel. The GMR and NL measurements could be repeated on 5 different lateral spin valves on the same chip with different gaps. However, we could not obtain a proper gap dependence of the spin signal to extract the spin diffusion length. For the quantitative analysis, Eq.~\ref{Hanle} is only valid when the resistance-area product $RA$ of the tunnel junction is much higher than the spin resistance of Ge $r_{Ge}$=$\rho\times l_{sf}$, $\rho$ and $l_{sf}$ being the Ge resistivity and the spin diffusion length. In our case, $RA$ is two orders of magnitude larger than $r_{Ge}$ which justifies the use of Eq.~\ref{Hanle}.\\  

\subsection*{Optical spin orientation and electrical spin detection}
The bulk Ge sample was first cleaned in acetone and isopropyl alcohol into an ultrasonic bath for 5 min and then rinsed into deionized water before being loaded into the MBE chamber. The native Ge oxide top layer was then thermally removed by annealing under ultrahigh vacuum to obtain a well-defined (2$\times$1) surface reconstruction. For the MTJ device, we first deposited a 8 nm-thick MgO layer at 310$^{\circ}$C followed by 10 minutes of annealing at 650$^{\circ}$C, 15 nm of Pt were then deposited on top at room temperature. Eight stripes of Pt/MgO with dimensions 1$\times$2 $\mu$m$^{2}$ and separated by 1 $\mu$m were then patterned by electron beam lithography and ion beam etching. Finally, the MTJ Pt(5nm)/Fe(15nm)/MgO(3.5nm) was grown at room temperature by electron beam evaporation and patterned using electron beam lithography and the lift-off process. For the ISHE device, starting from the same Ge surface, only a 15 nm-thick Pt layer was grown on Ge at room temperature. The 8 Pt stripes and the Pt stripe with dimensions 3$\times$1 $\mu$m$^{2}$ for the ISHE detection were then patterned by electron beam lithography and ion beam etching. For both the MTJ and ISHE devices, after passivating the surface with a 100 nm-thick SiO$_{2}$ layer, a Au(250nm)/Ti(10nm) stack was deposited to contact the MTJ and ISHE detector.\\
The sample is illuminated with a continuous wave (CW) laser diode working at a wavelength of 1550 nm ($h\nu$=0.8 eV) to be resonant with the direct bandgap of Ge. The light is circularly polarized using the combination of a polarizer (POL) rotated at 45$^{\circ}$ with respect to the neutral lines of a photoelastic modulator (PEM). The numerical aperture (NA) of the objective is 0.7 giving a full-width at half maximum beam size $w\approx 1.5\mu$m. The circular polarization is modulated at 50 kHz which allows for the synchronous detection of the electrical signal $\Delta V$ using a lockin amplifier. Finally, the optical image is recorded using a near infrared (NIR) InGaAs detector after the light crosses a beamsplitter (BS).\\
It is worth noticing that the detection pad for MTJ and ISHE measurements has been designed in order to avoid spurious electrical effects related to the electron diffusion in Pt. At variance from Ref. \onlinecite{Bottegoni2014}, where the ISHE detection was performed with a continuous thin Pt film, with a bar-shaped ISHE pad we do not detect any signal related to a component of the spin-polarization parallel to the Pt stripe edge. This indicates that the electromagnetic field modulation, operated by the Pt scatterers, generates only two complementary in-plane components of the spin polarization, perpendicular to the stripe edges.

\section*{Acknowledgements}

The authors acknowledge the financial support from the French National Research Agency through the ANR project SiGeSPIN \#ANR-13-BS10-0002. Partial funding is acknowledged to the CARIPLO project SEARCH-IV (grant 2013-0623). Dr. Edith Bellet-Amalric is also acknowledged for the x-ray diffraction analysis of GeOI.

\section*{Author contributions}

MJ and FB proposed and supervised the project. JW, FR, AM and MJ perfomed the MBE growths. CV processed the nanostructures. FR, CV and MJ performed the electrical measurements. AF, CZ, LG, CV, MC, MF, MJ and FB performed the optical measurements. HJ and JMG interpreted the electrical measurements using the lateral spin valves. JPA, GI and FC participated to the results discussion. All the authors contributed to the writing of the manuscript.

\section*{Competing financial interests}

The authors declare no competing financial interests.

\newpage

\begin{figure}[h!]
\caption{\textbf{Lateral spin valve fabricated on GeOI}. (a) RHEED patterns recorded along the [110] and [100] crystal axes of Ge at different stages of the epitaxial growth of the magnetic tunnel junction on Ge(100). (b) Sketch of the lateral spin valve used for non-local electrical spin injection and detection in $n$-Ge. The electrical current is applied between the hard magnetic layer and one ohmic contact in electron spin injection conditions. (c) Scanning electron microscopy image of the lateral spin valve. The magnetic field is applied at 45$^{\circ}$ from the injected spin direction.}
\label{Fig1}
\end{figure}

\begin{figure}[h!]
\caption{\textbf{Room temperature magnetoresistance measurements}. Three different measurement geometries are used: (a) local or GMR, (b) non-local and (c) Hanle. The applied DC current is 10 mA. The magnetic field is applied in-plane at 45$^{\circ}$ from the Fe electrode axis in (a) and (b) and perpendicular to the film plane in (c). The horizontal arrows indicate the field sweep directions. Orange (resp. black) is for increasing (resp. decreasing) magnetic field. All the curves have been shifted vertically so that the high field magnetoresistance $\Delta R$ is zero.}
\label{Fig2}
\end{figure}

\begin{figure}[h!]
\caption{\textbf{Setup and principle of optical spin orientation and electrical spin detection}. (a) Sketch of the confocal microscope setup used for optical spin orientation and electrical detection. (b) and (c) Right and left circularly polarized light impinging the Pt or Pt/MgO nanostructures and resulting in-plane spin polarization of electrons with opposite chirality at the two edges. (d) Voltage recorded between the MTJ and one ohmic contact while sweeping the laser spot along the $x$ axis. The white arrow indicates the magnetization direction of the MTJ. $V_{\sigma +}(x)$ (resp. $V_{\sigma -}(x)$) is for right (resp. left) circularly polarized light. $\Delta V(x)=V_{\sigma +}(x)-V_{\sigma -}(x)$ is the voltage recorded on the lock-in amplifier. Here, the distance between the MTJ and the Pt stripe is assumed to be much shorter than the spin diffusion length in germanium.}
\label{Fig3}
\end{figure}

\begin{figure}[h!]
\caption{\textbf{Lateral devices for optical spin generation and non-local electrical spin detection}. (a) and (b) Sketch of the MTJ and ISHE devices respectively. (c) and (d) Corresponding scanning electron microscopy images. For the MTJ device, the electrical contacts are taken between the MTJ and one ohmic contact made of Au(250nm)/Ti(10nm) directly grown on Ge. For the ISHE device, the contacts are directly taken on the Pt stripe.}
\label{Fig4}
\end{figure}

\begin{figure}[h!]
\caption{\textbf{Experimental evidence of optical spin generation and non-local electrical spin detection in Ge at room temperature}. (a) and (b) Optical images recorded on the MTJ and ISHE devices respectively using the confocal microscope. (c) and (d) Voltage signals simultaneously recorded using a lockin amplifier at the PEM modulation frequency. (e) and (f) Voltage intensity profiles along $x$ for $y$=0. $x$=0 is the detector position. Dots are experimental data and the solid lines correspond to the fits. The signals have been normalized to the laser power and expressed in $V/W$. As illustrated by the red dotted vertical line, the signal is zero at the center of the Pt stripe, positive (resp. negative) at the left (resp. right) edge.}
\label{Fig5}
\end{figure}

\clearpage

\includegraphics[width=\textwidth]{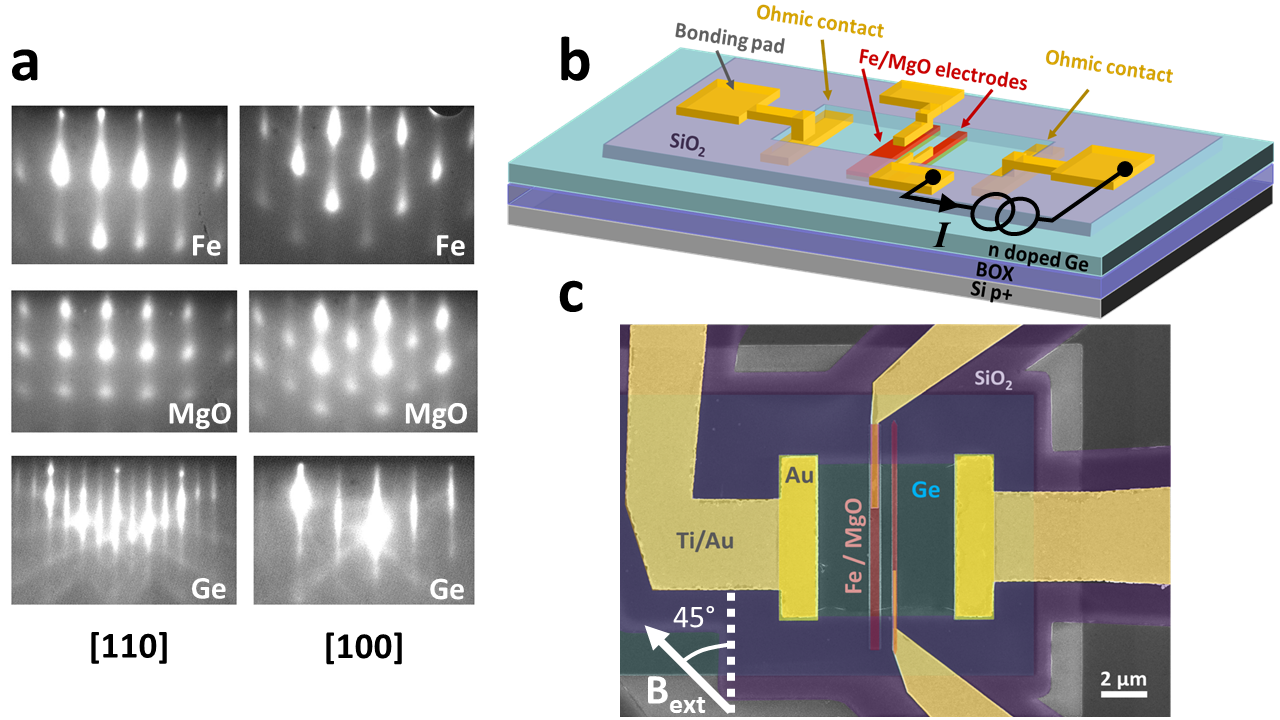}
\begin{center}
Figure 1
\end{center}

\clearpage

\includegraphics[width=\textwidth]{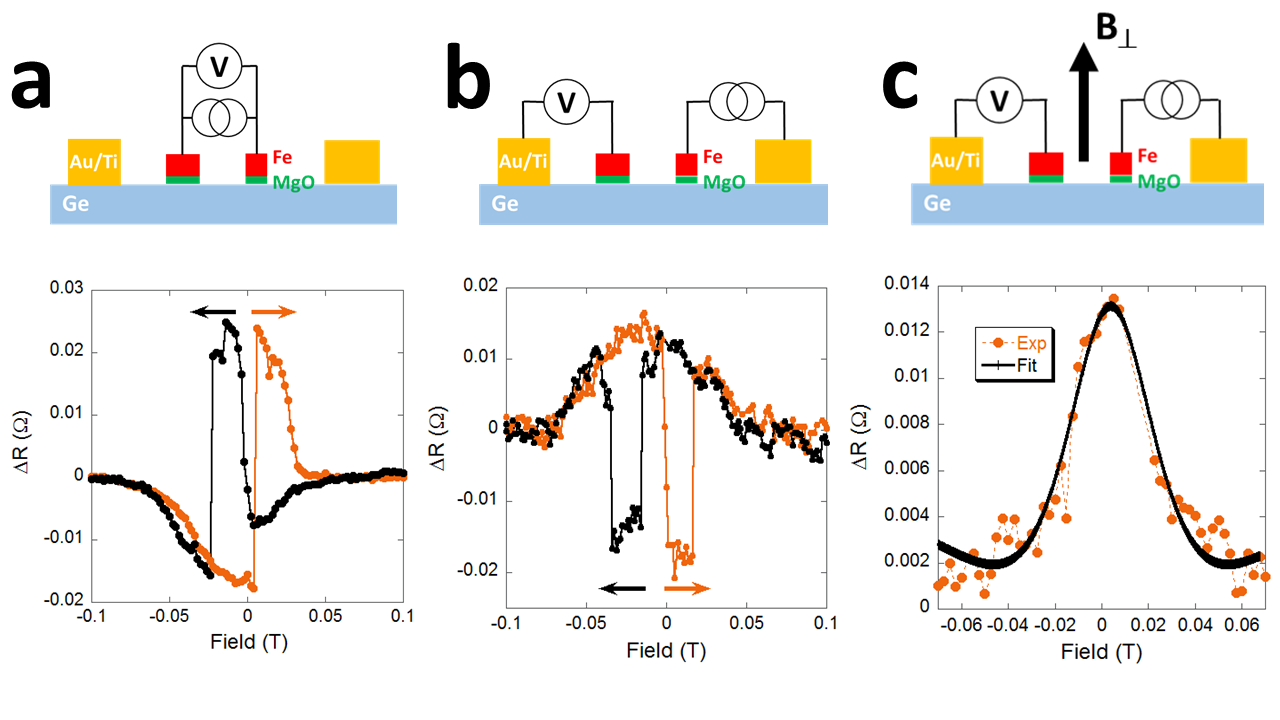}
\begin{center}
Figure 2
\end{center}

\clearpage

\includegraphics[width=\textwidth]{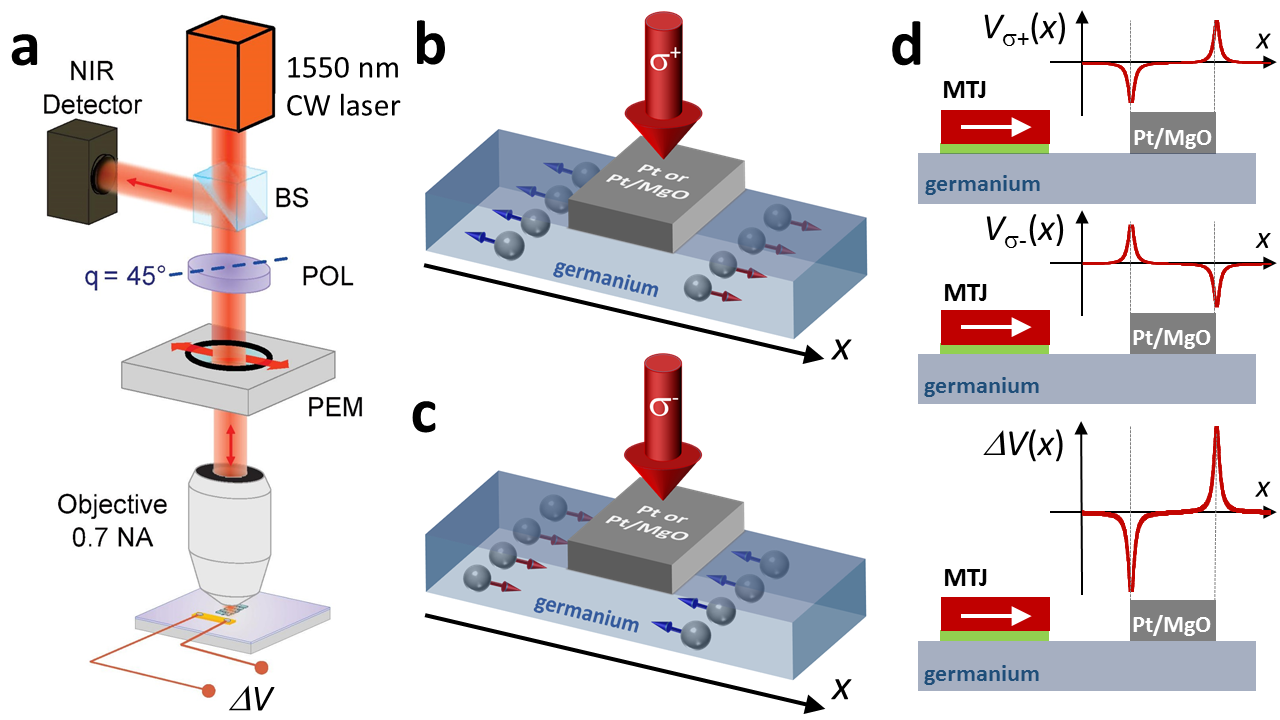}
\begin{center}
Figure 3
\end{center}

\clearpage

\includegraphics[width=\textwidth]{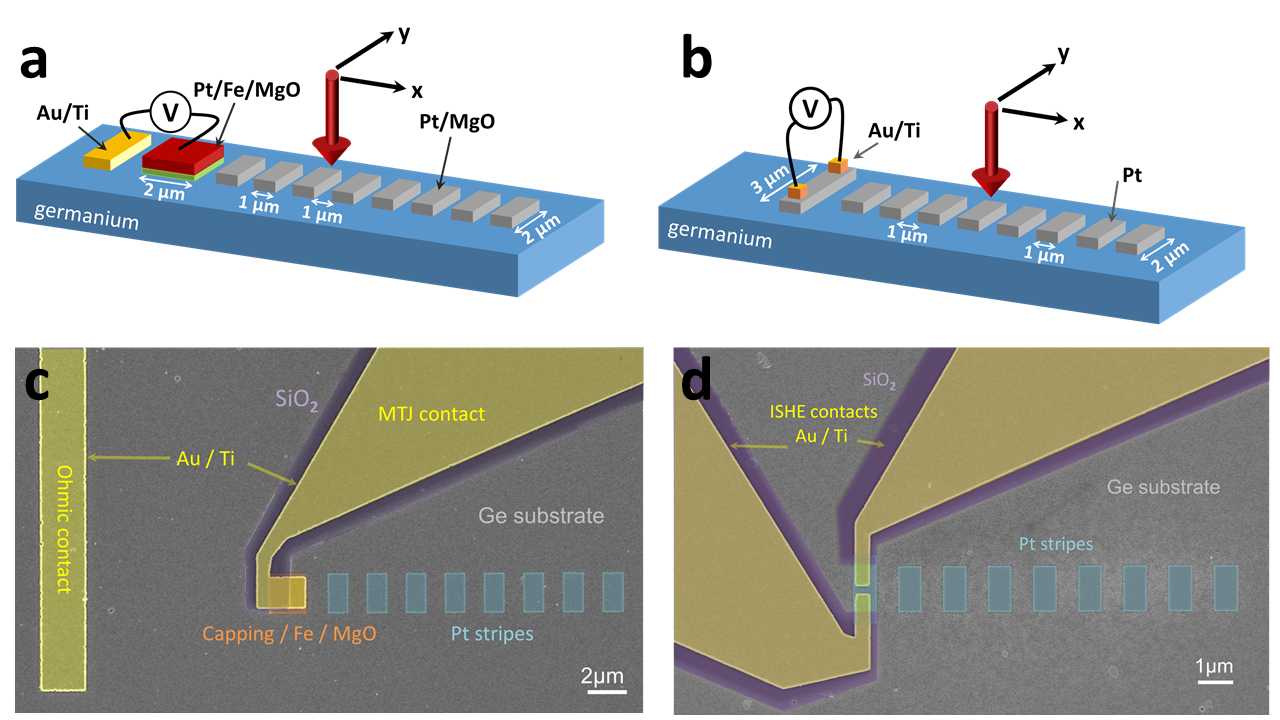}
\begin{center}
Figure 4
\end{center}

\clearpage

\includegraphics[width=\textwidth]{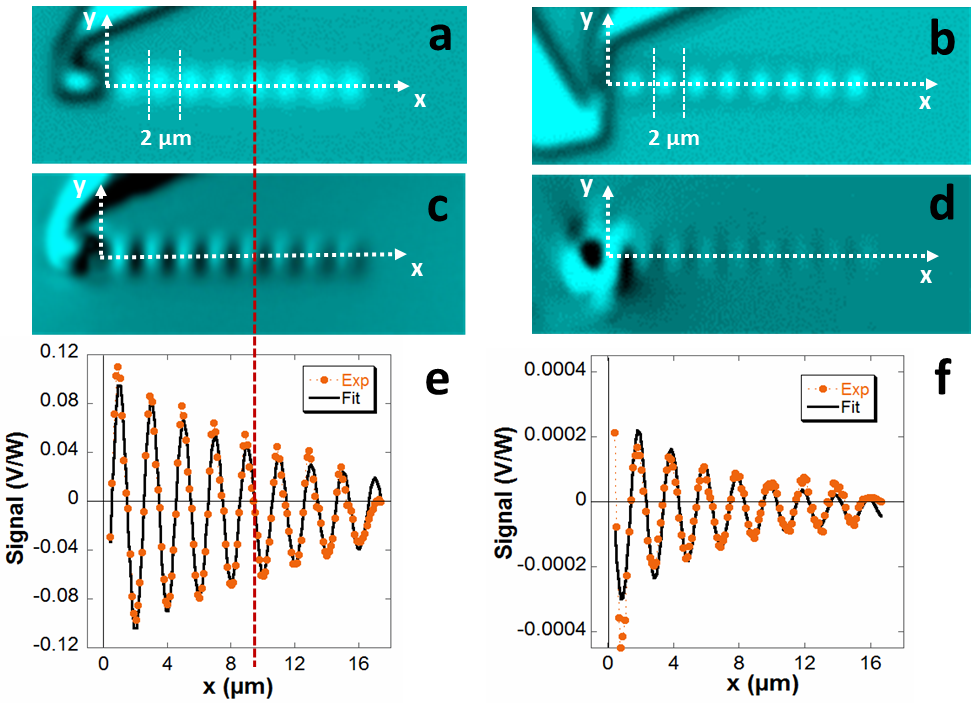}
\begin{center}
Figure 5
\end{center}

\end{document}